\documentclass{article}
\usepackage{spconf,amsmath,graphicx,hyperref,caption,rotating}
\usepackage{booktabs}
\usepackage{multirow}
\usepackage{cleveref}

\title{Can Audio Large Language Models Verify Speaker Identity?}
\name{Yiming Ren\textsuperscript{1}, Xuenan Xu\textsuperscript{1}, Baoxiang Li\textsuperscript{1}, Shuai Wang\textsuperscript{2}, Chao Zhang\textsuperscript{1}}
\address{\textsuperscript{1}Shanghai Artificial Intelligence Laboratory \\ \textsuperscript{2}Nanjing University}
\begin{document}
%
\maketitle
\begin{abstract}
This paper investigates adapting Audio Large Language Models (ALLMs) for speaker verification (SV).
We reformulate SV as an audio question–answering task and conduct comprehensive zero-shot evaluations on public benchmarks, showing that current ALLMs have limited zero-shot SV capability and often struggle in diverse acoustic conditions.
To address this challenge, we perform supervised fine-tuning on speaker verification data.
A rule-based hard pair sampling strategy is proposed to construct more challenging training pairs.
Lightweight fine-tuning substantially improves the performance, though there is still a gap between ALLMs and conventional models.
Then, we extend to text-dependent SV by jointly querying ALLMs to verify speaker identity and spoken content, yielding results competitive with cascaded ASR–SV systems.
Our findings demonstrate that with proper adaptation, ALLMs hold substantial potential as a unified model for robust speaker verification systems, while maintaining the general audio understanding capabilities.
\end{abstract}
\section{Introduction}
\label{sec:intro}

Speaker Verification (SV) is a critical technology for authenticating an individual's identity by analyzing biometric features within their voice signal.
SV plays a pivotal role in fields such as smart homes, personalized voice assistants, and secure access control.
With the rapid development of deep neural networks (DNN), popular solutions for SV typically consist of a DNN-based embedding extractor~\cite{snyder2018x,desplanques2020ecapa,zhou2021resnext}, followed by a scoring backend to determine speaker similarity.
These approaches have been shown to be highly effective under relatively simple testing scenarios~\cite{chen2024eres2netv2}.
However, under challenging real scenarios, such as noisy environments and device mismatch, the robustness of SV systems remains a challenge~\cite{al2021model}.


Motivated by the success of text-based Large Language Models (LLMs)~\cite{bai2023qwen}, Audio LLMs (ALLMs)~\cite{chu2023qwen,chu2024qwen2,ding2025kimi,tang2023salmonn,wu2025step} have recently witnessed rapid development.
Compared with conventional models where specific architectures are designed for different tasks, ALLMs achieve strong performance on diverse audio understanding tasks, spanning areas such as automatic speech recognition (ASR), audio captioning, and music question answering.
In addition to understanding tasks that involve processing audio inputs to produce textual outputs, recent ALLMs~\cite{ding2025kimi,wu2025step,xu2025qwen2} can generate both text and audio, supporting spoken dialogue between humans and ALLMs.

Due to their large parameter scale and the diversity of training tasks and data, ALLMs exhibit superior generalization capabilities compared to conventional task-specific smaller models.
For instance, they show some ability to recognize speaker-related attributes such as gender, age, and accent, suggesting a potential for performing SV tasks.
Nevertheless, current ALLM-based interactive systems remain largely insensitive to speaker identity in dialogue.
This naturally raises the question: to what extent can ALLMs effectively handle SV tasks?
To explore the potential of ALLMs for SV, we pioneer a comprehensive study addressing three key questions: (1) Can ALLMs perform zero-shot speaker verification? (2) How can we effectively adapt ALLMs for SV tasks? (3) Can ALLMs handle both text-independent and text-dependent SV scenarios?

First, we extensively evaluate ALLMs for SV in challenging conditions involving speaker pairs with similar attributes, such as gender, age, and recording device.
Evaluation pairs with specified conditions are constructed from public benchmark datasets: VoxCeleb~\cite{nagrani2017voxceleb}, CN-Celeb~\cite{fan2020cn} and 3D-Speaker~\cite{zheng20233d}.
To adapt ALLMs to Sv, the task is reformulated as an audio question answering (QA) task where two utterances, including enrollment and testing utterances, are fed to ALLMs.
We investigate four enroll-test prompting strategies and find that concatenation with silence achieves the best performance.
However, regardless of the evaluation condition or prompting strategy, ALLMs exhibit poor performance on SV.
This motivates our exploration of supervised fine-tuning approaches.
To expose ALLMs to more challenging pairs during training, we adopt a hard pair sampling strategy based on the aforementioned conditions (e.g., same gender or same age), rather than relying on random sampling.
Results show that lightweight fine-tuning substantially improves SV performance.
Remarkably, under the short-duration condition, fine-tuned ALLMs even surpass the strong conventional baseline, ECAPA-TDNN.

Finally, we extend our study to text-dependent SV, which requires simultaneous verification of both speaker identity and spoken content.
We reformulate text-dependent SV to an audio QA task where the model is required to judge whether two utterances are spoken by the same speaker and whether their textual content matches.
This formulation allows us to exploit the inherent speech understanding capabilities of ALLMs.
Experimental results demonstrate that ALLMs achieve competitive performance compared with cascaded approaches based on conventional models.

Our contribution can be summarized as follows:
\begin{itemize}
    \item To the best of our knowledge, this is the first work to prompt and adapt ALLMs for speaker verification. Comprehensive zero-shot evaluation reveals that current ALLMs exhibit limited performance on this task.
    \item We explore fine-tuning ALLMs for SV, highlighting the importance of the hard pair sampling strategy. The lightweight fine-tuning substantially enhances SV performance.
    \item We further extend ALLMs to text-dependent SV. The task reformulation of text-dependent SV leverages the understanding capabilities of ALLMs, achieving strong performance.
\end{itemize}

\section{Zero-Shot Evaluation of ALLMs on Speaker Verification}
\label{sec:zero_shot_allm_speaker}

\subsection{Speech Pair Prompting Strategy}
\label{subsec:speech_pair_prompt_strategy}
ALLMs are pre-trained with the Next Token Prediction (NTP) objective, so downstream audio understanding tasks need to be reformulated as text generation tasks to align with this paradigm.
For SV, we transform it to an audio-based QA task: given two utterances and a query,the model is expected to generate an answer determining whether the two utterances are from the same speaker.
Since both enrollment and test utterances are provided, different prompting strategies can be designed to organize the input pair:

\begin{enumerate}
  \item \textbf{Separate}: The two segments are input as independent utterances: ``Audio 1: \texttt{[audio1]}, Audio 2: \texttt{[audio2]}''. The textual prompt is ``Please determine whether the above two audio segments are from the same speaker or different speakers.''
  \item \textbf{Concat}: The two segments are concatenated into a single utterance: ``\texttt{[audio1]} $\parallel$ \texttt{[audio2]}''. The textual prompt is ``Please determine how many speakers are present in this audio segment.''.
  \item \textbf{Concat + Silence}: The two segments are concatenated with a 1-second silence interval inserted in between: ``\texttt{[audio1]} $\parallel$ \texttt{sil} $\parallel$ \texttt{[audio2]}''. The textual prompt is the same as \textbf{Concat}.
  \item \textbf{Mix}: The two segments are mixed and overlaid to form a single utterance: ``\texttt{[audio1$\oplus$audio2]}''. The textual prompt is ``This audio segment is composed of two audio tracks mixed together. Please determine whether these two audio tracks are from the same speaker or different speakers.''.
\end{enumerate}

Intuitively, the inserted silence between utterances explicitly separates the two segments, so we assume \textbf{Concat + Silence} may help ALLMs better differentiate speakers than \textbf{Concat}. 

\subsection{Evaluation Settings}
\label{subsec:eval_settings}
Three ALLMs are investigated: Qwen2-Audio (7B), Kimi-Audio (7B), and Step-Audio 2 (8B).
They all feed audio features with text token embeddings to LLMs for NTP.
Specifically, we use the following checkpoints: Qwen2-Audio-7B-Instruct, Kimi-Audio-7B-Instruct, Step-Audio-2-mini.

\begin{table}[ht]
\centering
\small
\caption{Details of challenge dimensions in the test set.}
\label{tab:test_set_details}
\begin{tabular}{lcc}
\toprule
\textbf{Dimension} & \textbf{Source} & \textbf{\# Pairs} \\
\midrule
Gender       & \multirow{3}{*}{VoxCeleb1~\cite{nagrani2017voxceleb}}     & 1500 \\
Language     &      & 1500 \\
Age          &      & 1500 \\
\midrule
Device       & \multirow{3}{*}{3D-Speaker-test~\cite{zheng20233d}}   & 1500 \\
Distance     &    & 1500 \\
Dialect      &    & 1500 \\
\midrule
Duration     & \multirow{2}{*}{CNCeleb-eval~\cite{fan2020cn}}     & 3000 \\
Scene     &      & 2000 \\
\bottomrule
\end{tabular}
\end{table}

Although there have been benchmarks such as VoxCeleb, we aim to evaluate model performance regarding specific challenging dimensions.
To this end, we construct a comprehensive test set by selectively sampling from existing benchmarks, with a focus on diverse dimensions.
For each dimension, positive and negative pairs are constructed by choosing pairs with the same or different corresponding attributes.
For example, positive pairs regarding age are from the same speaker, but with an age gap of over 10 years.
Negative pairs regarding gender are from different speakers, who share the same gender.
Detailed conditions are shown in \Cref{tab:test_set_details}.
A 1:1 ratio of positive and negative pairs is maintained across all dimensions.
Since ALLMs directly output text without similarity scores, we use accuracy for evaluation instead of the common Equal Error Rate (EER).


\begin{table*}[ht]
\centering
\caption{Zero-shot speaker verification performance of Audio Large Language Models across different prompting strategies and datasets. Accuracy (\%) is reported. Best results for each dimension are highlighted in bold. }
\label{tab:zeroshot_result}
\resizebox{\textwidth}{!}{
\begin{tabular}{@{} llccccccccccc@ {}}
\toprule
\multicolumn{2}{c}{Dataset} & \multicolumn{3}{c}{voxceleb} & \multicolumn{3}{c}{3d-speaker} & \multicolumn{5}{c}{cnceleb} \\
\cmidrule(lr){3-5} \cmidrule(lr){6-8} \cmidrule(lr){9-13}
\multicolumn{2}{c}{Dimension} & Gender & Lang & Age & Device & Dialect & Distance & Dur. $<$ 2s & Dur. 2-6s & Dur. $>$ 6s & Same Scene & Different Scene \\
\midrule
\multirow{3}{*}{Separate} & Kimi-Audio & \textbf{64.00} & \textbf{66.47} & \textbf{57.13} & 52.80 & 50.20 & 51.33 & \textbf{55.80} & 54.40 & 54.50 & \textbf{56.30} & 55.50 \\
& Qwen2-Audio & 53.40 & 52.00 & 57.67 & \textbf{55.13} & \textbf{51.73} & 50.13 & 51.50 & 51.20 & 49.40 & 49.90 & 50.00 \\
& Step-Audio2 & 52.73 & 51.80 & 51.67 & 51.33 & 50.53 & 50.00 & 50.00 & 50.00 & 50.00 & 50.00 & 50.00 \\
\midrule
\multirow{3}{*}{Concat} & Kimi-Audio & \textbf{68.00} & \textbf{65.27} & 58.47 & 52.93 & \textbf{58.33} & \textbf{53.53} & \textbf{57.00} & 63.20 & \textbf{74.30} & \textbf{64.80} & 54.00 \\
& Qwen2-Audio & 60.47 & 60.13 & 53.53 & 52.60 & 50.47 & 53.00 & 48.30 & 58.60 & 60.40 & 63.20 & 53.00 \\
& Step-Audio2 & 67.60 & 63.87 & 59.47 & \textbf{56.93} & 58.27 & 53.27 & 53.30 & \textbf{67.80} & 73.20 & 64.00 & \textbf{58.20} \\
\midrule
\multirow{3}{*}{Concat + Silence} & Kimi-Audio & \textbf{70.20} & \textbf{68.40} & \textbf{63.40} & 52.67 & 55.00 & 52.07 & 53.70 & 59.00 & \textbf{73.60} & 60.80 & 51.30 \\
& Qwen2-Audio & 59.40 & 58.60 & 53.87 & 52.20 & 50.40 & 51.00 & 50.60 & 59.90 & 59.10 & \textbf{63.20} & 55.00 \\
& Step-Audio2 & 64.20 & 60.40 & 56.80 & \textbf{57.47} & \textbf{58.07} & \textbf{55.33} & \textbf{54.60} & \textbf{67.80} & 71.80 & 62.90 & \textbf{59.10} \\
\midrule
\multirow{3}{*}{Mix} & Kimi-Audio & 50.00 & 50.00 & 50.00 & 50.07 & 49.93 & 49.87 & 50.00 & 50.20 & 50.10 & 49.80 & \textbf{49.90} \\
& Qwen2-Audio & \textbf{50.07} & \textbf{50.07} & \textbf{50.07} & \textbf{50.47} & \textbf{50.87} & 48.60 & \textbf{51.40} & \textbf{51.20} & \textbf{53.60} & \textbf{52.30} & 47.30 \\
& Step-Audio2 & 50.00 & 50.00 & 50.00 & 49.80 & 49.47 & 49.80 & 50.50 & 50.20 & 49.00 & 50.00 & 49.90 \\
\bottomrule
\end{tabular}
}
\end{table*}

\subsection{Zero-Shot Performance}

Zero-shot performance of ALLMs on SV is shown in \Cref{tab:zeroshot_result}.
First, comparison between prompting strategies reveals that the strategy is critical to leveraging pre-trained ALLMs for SV.
\textbf{Concat + Silence} achieves the best performance across most evaluation dimensions and ALLMs, which is consistent with our assumption.
\textbf{Mix} performs close to chance level, indicating that ALLMs are limited in understanding speech mixture.
Then, results across different challenge dimensions reveal that diverse acoustic conditions remain challenging for the robustness: although $\sim$70\% accuracy is achieved on long-duration utterances, the performance degrades markedly under challenging scenarios like cross-device and cross-dialect conditions.
Guided by the experimental results, we employ Kimi-Audio with the Concat + Silence strategy in the following fine-tuning stage.

\section{Supervised Fine-tuning}

\subsection{Fine-tuning Paradigm}
We construct the fine-tuning dataset \(\mathcal{D}_{\mathrm{ft}}\) by pairing each input $A_i$ with an instruction prompt \(q_i\) and a target label \(y_i\):
\[
\mathcal{D}_{\mathrm{ft}}=\{(A_i, q_i), y_i\}_{i=1}^{N}.
\]
Here, $A_i$ denotes one or more audio segments and $q$ denotes a templated-based prompt.
They are combined into a single prompt as an interleaved sequence of text and audio.

\paragraph*{Text-Independent (TI)} For TI data, we adopt the \textbf{Concat + Silence} strategy in \Cref{subsec:speech_pair_prompt_strategy} to form \((A_i, q)\).
The label $y_i \in \{\text{``one''},\text{``two''}\}$.

\paragraph*{Text-Dependent (TD)} For TD data, the input consists of an enrollment utterance indicating the target speaker, a test utterance, and a target text.
They are interleaved to form the prompt: ``Enrollment Audio: \texttt{[audio1]}, Test Audio: \texttt{[audio2]}. Whether the Test and Enrollment correspond to the same speaker and whether the Test matches a specified \texttt{[target text]}.''. The label $y_i$ is defined as a template-based textual response:
\[
y_i = \mathtt{``Speaker:\ Yes/No,\quad Content:\ Yes/No''}.
\]
The pre-trained ALLM \(\mathcal{M}\) is fine-tuned by minimizing the cross-entropy loss:
\[
\theta^{\ast}=\arg\min_{\theta}\sum_{i=1}^{N}\mathcal{L}_{\mathrm{CE}}\big(\mathcal{M}_{\theta}(A_i,q),y_i\big),
\]
where \(\theta\) denotes the trainable parameters.
During inference, corresponding places in the template are extracted from the answer to obtain TI or TD prediction results.

\subsection{Rule-based Hard Pair Sampling Strategy}
Since ALLMs are fine-tuned to differentiate whether speech pairs share the same speaker identity or textual content, the pair sampling strategy is vital to the discriminative capability of the model.
Inspired by the large zero-shot performance variations across different dimensions, we propose a rule-based hard pair sampling strategy, where challenging pairs are constructed with an explicit focus on different dimensions, as illustrated in \Cref{subsec:eval_settings}.


\begin{table*}[ht]
    \centering
    \setlength{\tabcolsep}{4pt} 
    \caption{
        Performance comparison between zero-shot and fine-tuned Kimi-Audio across diverse challenging conditions, including gender (Gen.), device (Dev.), dialect (Dial.), distance (Dist.), duration (Dura.), and scene (Sc.). The conventional ECAPA-TDNN is also involved for reference. The best and the second-best results are shown in bold and underlined text, respectively.
    }
    \small
    \label{tab:model_performance_no_percent}
    \begin{tabular}{@{}l|lccccccccccc@{}}
        \toprule
        \multicolumn{2}{c}{
        \multirow{2}{*}{\textbf{Model}} } & \multicolumn{3}{c}{\textbf{voxceleb}} & \multicolumn{3}{c}{\textbf{3d-speaker}} & \multicolumn{5}{c}{\textbf{cnceleb}} \\
        \cmidrule(lr){3-5} \cmidrule(lr){6-8} \cmidrule(lr){9-13}
        \multicolumn{1}{c}{} & & Gen. & Lang & Age & Dev. & Dial. & Dist. & Dur.$<$2s & Dur.2-6s & Dur.$>$6s & Same Sc. & Diff. Sc. \\
        \midrule
        \multirow{3}{*}{Kimi-Audio} & zero-shot & 70.20 & 68.40 & 63.40 & 52.67 & 55.00 & 52.07 & 53.70 & 59.00 & 73.60 & 60.80 & 51.30 \\
        & finetuning & \underline{95.07} & \underline{97.00} & \underline{92.40} & \underline{88.20} & \underline{89.00} & \underline{81.20} & \textbf{80.90} & \underline{85.80} & \underline{89.50} & \underline{85.10} & 77.70 \\
        & \hspace{1em} w. rand. sample  & 94.80 & 93.07 & 92.27 & 85.60 & 80.53 & 78.07 & 77.00 & 82.90 & 89.00 & 80.00 & \underline{79.00} \\
        \midrule
        \multicolumn{2}{c}{ECAPA-TDNN} & \textbf{99.33} & \textbf{99.27} & \textbf{94.13} & \textbf{94.67} & \textbf{93.00} & \textbf{88.27} & \underline{78.80} & \textbf{91.10} & \textbf{95.60} & \textbf{93.50} & \textbf{80.10} \\
        \bottomrule
    \end{tabular}
\end{table*}

\section{Experimental Setup}
\label{sec:exp_setup}
\subsection{Datasets}
For text-independent speaker verification, the ALLM is fine-tuned on the VoxCeleb2, CN-Celeb, and 3D-Speaker development sets (with 9 million training pairs in total), and its performance is evaluated on the corresponding test sets described in \Cref{subsec:eval_settings}.
For text-dependent tasks, we fine-tune and evaluate the model on the LibriSpeech~\cite{7178964} dataset, where positive and negative pairs are randomly constructed in equal proportions, yielding 280k training pairs and 5,560 evaluation pairs.

\subsection{Implementation Details}
We adopt Low-Rank Adaptation (LoRA)~\cite{hu2022lora} with the rank $r=16$ and scaling factor $\alpha = 32$, applied to all attention projection layers (excluding biases).
The training parameter number is 18M.
AdamW optimizer is used with the learning rate following a cosine schedule and an initial value of $10^{-5}$.



\section{Results}

\subsection{Fine-tuning Effects}

\Cref{tab:model_performance_no_percent} shows the performance of fine-tuned Kimi-Audio.
Compared with zero-shot results, fine-tuning significantly improves the verification performance, demonstrating that fine-tuning is crucial for unlocking the potential of ALLMs for SV.
With only 18M trainable parameters, the fine-tuned ALLM achieves $\sim$95\% accuracy on VoxCeleb gender and language evaluation scenarios. 
Since there is no modification to the model input format, fine-tuning can potentially retain the strong general audio understanding capabilities of the ALLM.

We further validate the effectiveness of our proposed rule-based hard pair sampling strategy by comparing it with random sampling, where both positive and negative pairs are randomly sampled.
Without rule-based sampling, performance degradation is observed in most testing conditions.
This indicates that hard sample mining is crucial to performance, which is also observed in prior metric learning studies~\cite{zhang2021understanding}.

Finally, we incorporate a widely adopted conventional model, ECAPA-TDNN, as a reference.
Although fine-tuning has substantially improved the performance of ALLM on SV, a notable gap still remains compared with the conventional method.
Interestingly, under challenging conditions where ECAPA-TDNN achieves about 80\% accuracy, the gap becomes considerably narrower.
In fact, ALLM even outperforms ECAPA-TDNN on short-duration conditions.
These findings suggest that ALLMs may exhibit stronger robustness in real-world noisy scenarios, while bridging the performance gap between ALLMs and conventional SV methods remains an important future direction.

\subsection{Extension to Text-Dependent SV}

\label{subsec:text_dependent_sv}

\begin{table}[t]
\centering
\caption{Model Performance on LibriSpeech-test (Text-Dependent SV).}
\label{tab:td_performance}
\setlength{\tabcolsep}{4pt} 
\renewcommand{\arraystretch}{1.1} 
\small 
\resizebox{\columnwidth}{!}{
\begin{tabular}{lccc}
\toprule
\textbf{Model} & \textbf{Spk Acc (\%)} & \textbf{Txt Acc (\%)} & \textbf{Acc (\%)} \\
\midrule
Kimi-Audio (zero-shot)    & 62.09  & 89.61  & 52.31 \\
Kimi-Audio (finetuning)    & 98.92  & \textbf{99.95}  & \textbf{98.87} \\
Whisper + ECAPA-TDNN & \textbf{99.08}  & 99.75  & 98.83 \\
\bottomrule
\end{tabular}
}
\end{table}

The results of adapting ALLMs for text-dependent SV are presented in \Cref{tab:td_performance}.
We compare ALLMs with a cascaded approach where Whisper~\cite{radford2023robust} is used for text recognition and ECAPA-TDNN for speaker verification.
Consistent with the results in \Cref{tab:zeroshot_result}, ALLM performs poorly on SV, while maintaining a high accuracy of text verification.
This is expected as ASR is incorporated in ALLM pre-training.
ALLM demonstrates strong performance after fine-tuning, slightly surpassing the cascaded approach in overall accuracy while achieving comparable speaker accuracy.
These results suggest that ALLMs have substantial potential to serve as a unified SV system.
Unlike conventional SV methods that focus solely on speaker-related features, ALLMs inherently capture semantic content.
Consequently, ALLMs can be endowed with speaker identification capabilities without compromising their original general audio understanding performance.

\section{Conclusion}
\label{sec:format}
This study pioneers the use of ALLMs for speaker verification.
Zero-shot evaluation shows near-chance performance in challenging scenarios. Lightweight fine-tuning via LoRA significantly enhances the performance, where the rule-based hard pair sampling strategy is critical.
Extending to text-dependent SV leverages ALLMs' intrinsic speech understanding capabilities and achieves results comparable to cascaded conventional ASR-SV systems.
Future work will focus on bridging the gap between ALLMs and specialized conventional models on SV performance.

\bibliographystyle{IEEEtran}
\bibliography{strings,refs}

\end{document}